\documentclass{PoS}
\usepackage{epsf}

\newcommand{\half}{\mbox{\small $\frac{1}{2}$}}          
\newcommand{\imp}{\mbox{\tiny $I\!M\!P$}}                

\def\lsim{\mathrel{\rlap{\lower4pt\hbox{\hskip1pt$\sim$}}
    \raise1pt\hbox{$<$}}}                
\def\gsim{\mathrel{\rlap{\lower4pt\hbox{\hskip1pt$\sim$}}
    \raise1pt\hbox{$>$}}}                
\title{
\begin{picture}(0,0)(0,0)%
   \put(0,130){\makebox(0,0)[l]{\textnormal{\normalsize
                  DESY 06-169, Edinburgh 2006/31, Liverpool LTH 724}}}%
\end{picture}%
       Simulating at Realistic Quark Masses:
             Pseudoscalar Decay Constants and Chiral Logarithms}

\ShortTitle{Simulating at Realistic Quark Masses $\ldots$}

\author{Meinulf G\"ockeler$^{a}$, \speaker{Roger Horsley}$^b$,
        Yoshifumi Nakamura$^c$, Dirk Pleiter$^c$,
        Paul E.~L. Rakow$^d$, Gerrit Schierholz$^{ce}$,
        Wolfram Schroers$^c$, Thomas Streuer$^f$, Hinnerk St\"uben$^g$
        and James M. Zanotti$^b$ \\
        \llap{$^a$} Institut f\"ur Theoretische Physik,
                    Universit\"at Regensburg, \\
                    93040 Regensburg, Germany \\
        \llap{$^b$} School of Physics, University of Edinburgh, \\
                    Edinburgh EH9 3JZ, UK \\
        \llap{$^c$} John von Neumann Institute NIC / DESY Zeuthen, \\
                    15738 Zeuthen, Germany \\
        \llap{$^d$} Department of Mathematical Sciences,
                    University of Liverpool, \\
                    Liverpool L69 3BX, UK \\
        \llap{$^e$} Deutsches Elektronen-Synchrotron DESY, \\
                    22603 Hamburg, Germany \\
        \llap{$^f$} Department of Physics and Astronomy,
                    University of Kentucky, \\
                    Lexington KY 40506, USA \\
        \llap{$^g$} Konrad-Zuse-Zentrum f\"ur Informationstechnik Berlin, \\
                    14195 Berlin, Germany \\
        E-mail: \email{meinulf.goeckeler@physik.uni-regensburg.de},
                \email{rhorsley@ph.ed.ac.uk},
                \email{yoshifumi.nakamura@desy.de},
                \email{dirk.pleiter@desy.de},
                \email{rakow@amtp.liv.ac.uk},
                \email{gsch@mail.desy.de},
                \email{wolfram.schroers@feldtheorie.de},
                \email{thomas.streuer@uky.edu},
                \email{stueben@zib.de},
                \email{jzanotti@ph.ed.ac.uk}  }

\author{QCDSF--UKQCD Collaboration}

\abstract{Due to improvements in computer performance and algorithms,
          the rapidly increasing cost for unquenched Wilson-type fermions
          with lighter quarks has been ameliorated and new simulations
          are now possible. Here we present results using two flavours
          of $O(a)$-improved Wilson fermions for meson decay constants
          at pseudoscalar masses down to 320\,MeV. Results are at
          several lattice spacings down to about 0.07\,fm and include
          a non-perturbative determination of the renormalisation constant.
          This enables us to attempt contact with (partially quenched)
          chiral perturbation theory.}

\FullConference{XXIVth International Symposium on Lattice Field Theory \\
                July 23-28, 2006 \\
                Tucson, Arizona, USA}

\begin{document}


\section{Introduction}

Chiral extrapolations of lattice data to the physical pion mass
and the continuum or $a\to 0$ limit remain major sources of
systematic uncertainty in the determination of hadron
masses and matrix elements. A test that lattice QCD must 
successfully pass before predictions can be fully trusted is to reproduce 
known experimental results. One such indicator is the determination
of meson decay constants, such as $f_{\pi^+}$ and $f_{K^+}$, with 
phenomenological values of $92.42 \pm 0.07 \pm 0.25 \,\mbox{MeV}$ and
$113.0 \pm 1.0 \pm 0.3 \,\mbox{MeV}$, \cite{yao06} respectively.
The problem is that simulations for smaller quark masses rapidly become
very costly in computer time. Recent advances have been on two fronts:
firstly faster machines have become available, with speeds in the Tflop
range and secondly the hybrid Monte Carlo algorithm used in the
simulations has been improved. In particular in the new simulations
reported here, we have used trajectory length one with
three time scales in the molecular dynamic
step (one for the glue term \cite{alikhan03a} and now two \cite{urbach05a}
for the fermion term in the action) which allowed the computationally
expensive pieces to be updated less frequently. This was coupled with
the use of an auxiliary fermion mass, \cite{hasenbusch01}.

The results reported here use Wilson glue (plaquette) and two mass
degenerate $O(a)$-improved Wilson quarks (so effectively we are
simulating 2-flavour QCD). As emphasised by L\"uscher \cite{luscher05a},
these `clover' fermions are well understood: in particular the
addition of certain irrelevant terms, both in the action
and operators, and the non-perturbative determination of their coefficients
allow discretisation errors to be reduced to $O(a^2)$. For example
adding the `clover' term together with the appropriate coefficient
$c_{sw}$ is sufficient to determine the $O(a)$-improved masses,
such as the pseudoscalar mass $m_{ps}$ while to determine the
decay constant, given by
\begin{eqnarray}
   \langle 0| {\cal A}_4 |ps \rangle = {f_{ps}\over \sqrt{2}} m_{ps} \,,
\end{eqnarray}
the axial current $\cal A_\mu$ must also be $O(a)$-improved, which
can be achieved by setting
\begin{eqnarray}
    {\cal A}_\mu = Z_A {\cal A}_\mu^{\imp}\,, \qquad
    {\cal A}_\mu^{\imp} = \left( 1 + \half b_A( am_{q_1} + am_{q_2}) \right)
                             \left( A_\mu + c_A a\partial_\mu P \right) \,,
\label{Aimp}
\end{eqnarray}
where $A_\mu = \overline{q}_1\gamma_\mu\gamma_5 q_2$ and
$P = \overline{q}_1\gamma_5 q_2$.

\begin{figure}[th]
   \hspace{1.00in}
   \epsfxsize=8.00cm
      \epsfbox{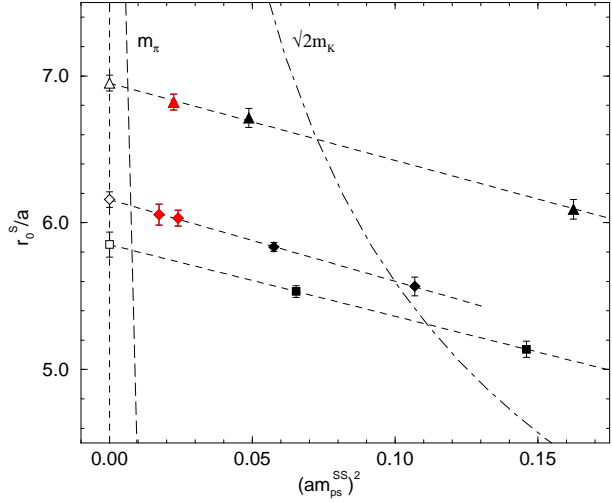}
   \caption{Results for $r_0^S/a$ versus $(am_{ps}^{SS})^2$
            for $\beta = 5.25$ (squares), $5.29$ (diamonds)
            and $5.40$ (triangles). The new runs are shown in red.
            Linear fits have been used to
            extrapolate $r^S_0/a$ to the chiral limit,
            the results being denoted by open symbols.
            The vertical dashed lines (left to right) represent
            the chiral limit, and using LO $\chi$PT approximate
            positions of a $\bar{l}l$, and fictitious $\bar{s}s$
            pseudoscalar particle computed from $\pi^+$ and $K^+$
            respectively.}
   \label{fig_amps2_r0oa_nf2_lin_060927_fine}
\end{figure}

Several years ago we started simulations at four $\beta$ values
and reached pseudoscalar masses of $\sim 600\,\mbox{MeV}$. We have started
new simulations at $\beta = 5.29$ and $\beta = 5.40$ at lower quark masses.
Our present status of the lower quark mass runs used in this
report is given in table~\ref{table_status}.
\begin{table}[th]
   \begin{center}
      \begin{tabular}{||l|l|l|c||c|c|l|l|c||}
         \hline
          $\beta$ & $\kappa^S$ & Volume & Trajs &
          $m^{SS}_{ps}/m^{SS}_{vec}$ & $m^{SS}_{ps}L$ & $a[\mbox{fm}]$ 
              & $L[\mbox{fm}]$ & $m^{SS}_{ps}[\mbox{MeV}]$ \\
         \hline
 5.25 & 0.13575& $24^3\times 48$ & 6000 &
 0.60 & 6.1    & 0.085 & 2.05    & 590 \\
         \hline
 5.29 & 0.1359 & $24^3\times 48$ & 4900 &
 0.61 & 5.8    & 0.081 & 1.95    & 580 \\
 5.29 & 0.1362 & $24^3\times 48$ & 3400 &
 0.42 & 3.7    & 0.081 & 1.95    & 380 \\
 5.29 & 0.13632& $32^3\times 64$ & 1200 &
 0.42 & 4.2    & 0.081 & 2.60    & 320 \\
         \hline
 5.40 & 0.1361 & $24^3\times 48$ & 3600 &
 0.63 & 5.3    & 0.072 & 1.73    & 610 \\
 5.40 & 0.1364 & $24^3\times 48$ & 2800 &
 0.51 & 3.6    & 0.072 & 1.73    & 410 \\
         \hline
      \end{tabular}
   \end{center}
\caption{Present data sets. The new runs are at $(\beta, \kappa^S) =$
         $(5.29,0.1362)$, $(5.29,0.13632)$ and $(5.40,0.1364)$, where
         `$S$' means sea quark. $m^{SS}_{ps}$, $m^{SS}_{vec}$ are the
         pseudoscalar and vector particle masses respectively.
         $L$ is the box size. For comparison, experimentally
         $m_{\pi^+}/m_{\rho^+} \sim 0.18$ and
         $m_{\pi^+} \sim 140\,\mbox{MeV}$.}
\label{table_status}
\end{table}
This has enabled us to reach pseudoscalar masses of $350\,\mbox{MeV}$
or less. The force-scale was the unit used to set the scale, 
together with a reference value $r_0 = 0.5\, \mbox{fm}$. Our results
for $r^S_0/a$ are shown in fig.~\ref{fig_amps2_r0oa_nf2_lin_060927_fine}.
In our extrapolations
we presently include results for heavier pseudoscalar masses;
hopefully the situation will improve with more smaller quark mass
results being generated so that a linear fit for the lighter quark
masses will suffice. The extrapolated values of $r^S_0/a$ in the
chiral limit $(r_{0}/a)_c$ are used to determine the scale.


\section{Chiral perturbation theory}


While the sea quark masses ($S$) are given (implicitly) in
table~\ref{table_status}, valence quarks ($V$) do not have to be chosen
to have the same mass. Chiral Perturbation Theory, $\chi PT$, has been 
extended to Partially Quenched Chiral Perturbation Theory, $PQ\chi PT$,
\cite{bernard93a,sharpe97a}. While it is expensive to generate dynamical
configurations, it is computationally cheaper to evaluate
correlation functions on these configurations, so that a range
of valence quark masses can be used. Using the Leading Order, LO, and
Next to Leading Order, NLO, results
\cite{sharpe97a} for the pseudoscalar masses and decay constants in
terms of the quark mass, we eliminate (iteratively) the quark mass
from these equations to give for degenerate mass valence quarks
\begin{eqnarray}
   F_{ps}^{VV}
      = f_a + f_b (M_{ps}^{SS})^2 + f_c (M_{ps}^{VV})^2 
              + f_d \left( (M_{ps}^{SS})^2 + (M_{ps}^{VV})^2 \right)
                    \ln \left( (M_{ps}^{SS})^2 + (M_{ps}^{VV})^2 \right) \,,
\label{fpsVV}
\end{eqnarray}
where we have also rescaled the pseudoscalar mass, $m_{ps}^{AB}$, and
decay constant, $f_{ps}^{AB}$ with say $r_{0c}$, ie
\begin{equation}
   M_{ps}^{AB} = r_{0c} m_{ps}^{AB}\,, \qquad
   F_{ps}^{AB} = r_{0c} f_{ps}^{AB}\,, \qquad
\end{equation}
with $A, B \in \{ V, S \}$. $f_a$ (the LO result) and $f_i$, $i = b, c, d$
are given in terms of the low energy constants, LECs,
$\alpha_4 \sim -0.76$, $\alpha_5 \sim 0.5$ (evaluated at
a scale $\mu = \Lambda_\chi = 4\pi f_0$) and $f_0 \sim 86.2\,\mbox{MeV}$
(the decay constant in the chiral limit) \cite{colangelo03a} by%
\footnote{If we had rescaled the pseudoscalar mass and decay constant with
$r^S_0/a$ (rather than $(r_0/a)_c$ as here)
$r ^S_0 / a  = ( r_0 / a )_c
                \left( 1 - r_m (M_{ps}^{SS})^2 + \ldots \right)$
would just give an additional term $-(4\pi f_0r_{0c})^2 r_m$ in
eq.~(\ref{fi_defs}) in the square brackets for $f_b$.}
\begin{eqnarray}
   f_a &=& r_{0c} f_0
                                           \nonumber \\
   f_b &=& { 1 \over (4\pi)^2 r_{0c}f_0 }
           \left[ {1\over 2} n_f \alpha_4
                  + {1\over 4}n_f \ln \left( 2 (4\pi r_{0c}f_0)^2 \right)
           \right]
                                           \nonumber \\
   f_c &=&  { 1 \over (4\pi)^2 r_{0c}f_0 }
           \left[ {1\over 2} \alpha_5
                  + {1\over 4}n_f \ln \left( 2 (4\pi r_{0c}f_0)^2 \right)
           \right]
                                           \nonumber \\
   f_d &=& - { 1 \over (4\pi)^2 r_{0c}f_0 } {1 \over 4} n_f \,.
\label{fi_defs}
\end{eqnarray}
When $V = S$, eq.~(\ref{fpsVV}) further simplifies to
\begin{eqnarray}
   F_{ps}^{SS} = f_a + (f_b + f_c + 2f_d \ln 2 ) (M_{ps}^{SS})^2 
                   + 2f_d (M_{ps}^{SS})^2 \ln (M_{ps}^{SS})^2 \,.
\label{fpsSS}
\end{eqnarray}

From eq.~(\ref{fpsVV}) the pion and kaon decay constants
can be found. We have two mass degenerate sea quarks
which we associate with the light quark ($l$ where
$m_l = ( m_u + m_d )/2$), together with two valence quarks,
which we associate with either the light, $l$, quark or the strange, $s$,
quark. Again manipulating the structural form of the LO and NLO equations
gives the result
\begin{eqnarray}
   F_{\pi^+} &=& f_a + (f_b+f_c+2f_d \ln 2) M_{\pi^+}^2
                     + 2 f_d M_{\pi^+}^2 \ln M_{\pi^+}^2
                                           \label{Fpi_formula} \\
   F_{K^+}   &=& f_a + \left( f_b + f_d \left(\ln 2 + {2\over n_f^2}\right)
                       \right) M_{\pi^+}^2
                     + \left( f_c + f_d \left( \ln 2 - {2\over n_f^2} \right)
                       \right) M_{K^+}^2
                                           \nonumber \\
             & & + f_d \left( 1 - {1\over n_f^2}\right)
                      M_{\pi^+}^2 \ln M_{\pi^+}^2
                 + { f_d \over n_f^2 } M_{\pi^+}^2
                       \ln \left( 2M_{K^+}^2 - M_{\pi^+}^2 \right)
                 + f_d M_{K^+}^2 \ln M_{K^+}^2 \,.
                                           \label{FK_formula}
\end{eqnarray}
Determining the $f_a$ and $f_i$, $i = b, c, d$ coefficients means that
the pion and kaon decay constants can be found. While degenerate
quark masses are sufficient, see eq.~(\ref{fpsVV}), for both
pion and kaon decay constants, only the pion decay constant is
possible with just sea quarks, eq.~(\ref{fpsSS}).

Detecting chiral logarithms is a notorious problem, but is necessary
as it shows that we are entering a regime where $\chi$PT is valid.
This is particularly difficult for decay constants, as can be
seen from eq.~(\ref{fpsVV}) that this term is
$\propto \left( (M_{ps}^{SS})^2 + (M_{ps}^{VV})^2 \right)
                   \ln \left( (M_{ps}^{SS})^2 + (M_{ps}^{VV})^2 \right)$
which for fixed $(M_{ps}^{SS})^2$ does not vary much with $(M_{ps}^{VV})^2$.
We wish for a term $\propto (M_{ps}^{SS})^2 \ln (M_{ps}^{VV})^2$.
As suggested in \cite{sharpe97a} considering the ratio
\begin{eqnarray}
   R \equiv { F^{VS}_{ps} \over \sqrt{ F^{VV}_{ps} F^{SS}_{ps} } } - 1
      =  { f_d \over n_f^2 f_a } \left( (M_{ps}^{SS})^2 \ln
                          { (M_{ps}^{VV})^2 \over (M_{ps}^{SS})^2 } 
                           + (M_{ps}^{SS})^2 - (M_{ps}^{VV})^2 
        \right) \,,
\label{enhanced_log}
\end{eqnarray}
(with $f_d / (n_f^2 f_a) = - 1 / ( 4n_f (4\pi r_{0c}f_0)^2)$
enhances these chiral logarithms. The disadvantage is that
mixed quark mass correlators ($V \not= S$) must be computed. Note
also that eqs.~(\ref{fpsVV}), (\ref{enhanced_log}) probe different
parts of the $\chi$PT expression; as can be seen from eqs.~(19) and (20)
of \cite{sharpe97a}, eq.~(\ref{enhanced_log}) sees only the $O(1/n_f)$
terms, while eq.~(\ref{fpsVV}) probes the remaining $O(1)$, $O(n_f)$ terms.


\section{Results}


We use the well-established procedure outlined in \cite{gockeler97a}
to compute decay constants. We only note here that in eq.~(\ref{Aimp}),
the improvement coefficient, $c_A$, has been computed non-perturbatively,
\cite{dellamorte05a}, while $b_A$ is only known perturbatively
(we use a tadpole improved version here, \cite{alikhan06a}).
We expect, however, that as the quark masses used here are quite small
this leads to negligible corrections. The renormalisation constant has
also been non-perturbatively computed, \cite{dellamorte05a,alikhan06a}
(the differences between these results appear to be $O(a^2)$ and hence
vanish in the continuum limit).

\begin{figure}[th]
   \hspace{1.00in}
   \epsfxsize=8.00cm
      \epsfbox{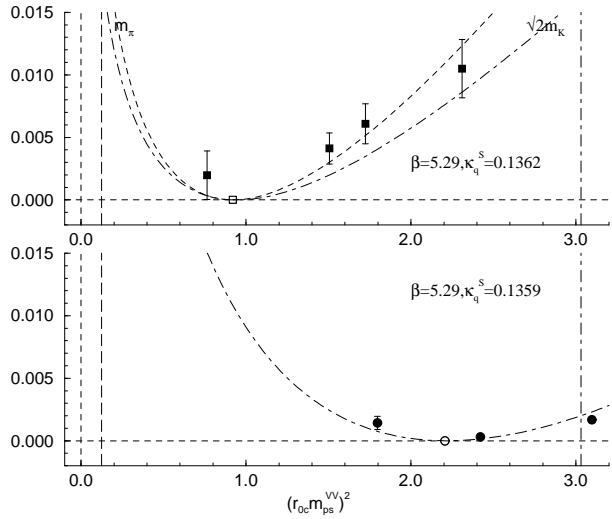}
   \caption{Results for $R$, eq.~(\protect\ref{enhanced_log}) for
            $(\beta, \kappa^S) = (5.29, 0.1359)$ (circles) and
            $(5.29, 0.1362)$ (squares), against
            $(M_{ps}^{VV})^2 \equiv (r_{0c}m^{VV}_{ps})^2$.
            The opaque symbols represent the points where $V \equiv S$
            when $R \equiv 0$ identically.
            The dash-dotted curves are also defined in
            eq.~(\protect\ref{enhanced_log}) and are plotted
            with $c = - 0.01659$ (using $r_0f_0 \sim 0.218$).
            The dashed curve is a fit, yielding
            $c \sim - 0.0227$ or $r_{0c}f_0 \sim 0.187$. Other notation
            as fig.~\protect\ref{fig_amps2_r0oa_nf2_lin_060927_fine}.}
   \label{fig_r0cmps2_fpsrat_b5p29_2pic_060923}
\end{figure}

We first investigate to see if we are entering a region where chiral
logarithms are becoming visible.
In fig.~\ref{fig_r0cmps2_fpsrat_b5p29_2pic_060923} we show $R$
defined in eq.~(\ref{enhanced_log}) for $\beta =5.29$ and
$\kappa = 0.1359$ and $0.1362$, together with the curve also given in 
eq.~(\ref{enhanced_log}). While we do not expect much influence from the
chiral logarithm, the curves track the data quite well, indeed out
to reasonably large quark masses. So it would appear the chiral logarithms
are visible of about the expected size. (But note the $y$-axis
scale -- we have subtracted $1$, so really this is a very small effect
of $O(1\%)$.)

To determine the decay constants we must first take the continuum
limit of the data, and then determine $f_a$ and $f_i$, $i = b, c, d$.
But as $\chi$PT is an infra-red expansion, while the $a^2 \to 0$ limit
is ultra-violet and as we are using $O(a)$-improved fermions then we
expect that there will be no problems with the order of the limits, ie
first chiral and then continuum. More drastically we shall presently assume
that we can ignore any $O(a^2)$ error. This assumption must however
be checked in the future.

\begin{figure}[th]
   \hspace{1.00in}
   \epsfxsize=8.00cm
      \epsfbox{
            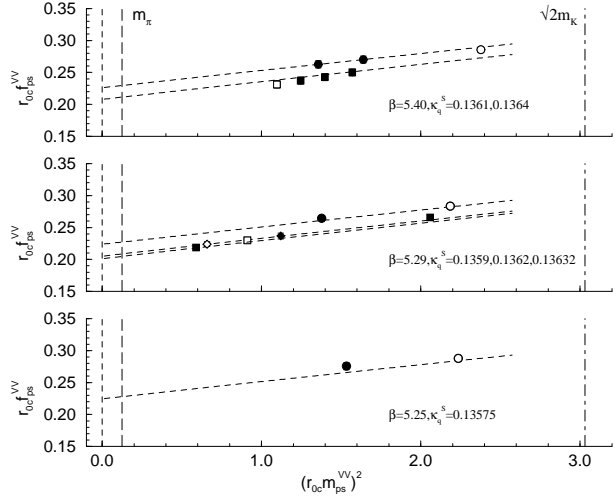}
   \caption{Results for $F_{ps}^{VV} \equiv r_{0c}f_{ps}^{VV}$ for
            the data sets given in table~\protect\ref{table_status}.
            The fit curves are given by eq.~(\protect\ref{fpsVV}),
            for a common parameter set. The opaque symbols represent
            the points where $V \equiv S$. Other notation as
            fig.~\protect\ref{fig_amps2_r0oa_nf2_lin_060927_fine}.}
   \label{fig_r0cmps2_r0cfps_b5p25+b5p29+b5p40_3pic_c2p20_060927}
\end{figure}

We now turn to a consideration of the partially quenched results.
In fig.~\ref{fig_r0cmps2_r0cfps_b5p25+b5p29+b5p40_3pic_c2p20_060927}
we show the results together with a fit from eq.~(\ref{fpsVV}).
This is a global fit giving one parameter set $f_a$ and $f_i$, $i = b, c, d$
with values $0.190(16)$, $0.017(13)$, $0.030(12)$, $-0.0015(59)$ 
respectively. (The fit is reasonably good given the fact that the data
has three varying parameters: $\beta$, $\kappa^S$ and $\kappa^V$.)
We first note the the value of $f_a \equiv r_{0c}f_0$ is in good 
agreement with the value determined from $R$. However using this
value to determine $f_d$ (see eq.~(\ref{fi_defs})) gives $\sim -0.017$,
indicating that we should be seeing a much stronger logarithmic dependence
(indeed the curves are almost linear). Furthermore using $f_a$ and
$f_b$, $f_c$ gives from eq.~(\ref{fi_defs}) the values
$\alpha_4 \sim -0.71$, $\alpha_5 \sim -0.63$. $\alpha_4$ is in
reasonable agreement with other phenomenological estimates;
but $\alpha_5$ is not. So at the moment there is no unambiguous
confirmation of $\chi$PT.

\begin{figure}[th]
   \hspace{1.00in}
   \epsfxsize=8.00cm
      \epsfbox{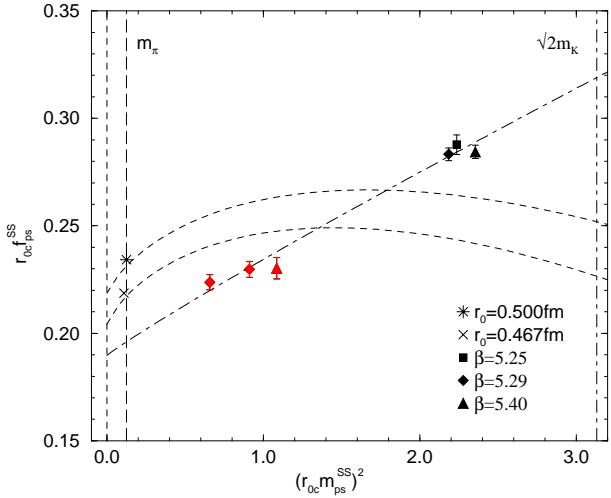}
   \caption{Results for $F_{ps}^{SS} \equiv r_{0c}m_{ps}^{SS}$
            for the data sets in table~\protect\ref{table_status}.
            The fit curve (dashed-dotted line) is taken from
            eq.~(\protect\ref{fpsSS}) using the parameters that
            have been determined from
     fig.~\protect\ref{fig_r0cmps2_r0cfps_b5p25+b5p29+b5p40_3pic_c2p20_060927}.
            The dashed curves are possible phenomenological curves,
            using eqs.~(\protect\ref{fi_defs}) and (\protect\ref{fpsSS}).
            The experimental value of $F_{\pi^+} \equiv r_0 f_{\pi^+}$
            is indicated by a star and cross for
            $r_0 = 0.5\,\mbox{fm}$ and $0.467\,\mbox{fm}$ respectively.
            Other notation as
            fig.~\protect\ref{fig_amps2_r0oa_nf2_lin_060927_fine}.}
   \label{fig_r0cmps2_r0cfps_nf2_060927_lat06}
\end{figure}

Consider now the sea quarks alone. In
fig.~\ref{fig_r0cmps2_r0cfps_nf2_060927_lat06}
we show these results. The curve joining the points uses the previously
determined $f_a$, $f_i$, $i = b, c, d$ coefficients. Consistency is
seen. The position of the new results is perhaps surprising because
they have dropped to almost below where the phenomenological value
might lie. Also shown is a possible phenomenological curve from 
eq.~(\ref{fi_defs}) for $r_0 = 0.5\,\mbox{fm}$. As previously found here
there is little agreement with the lattice results. Reducing the $r_0$
scale helps somewhat, the second curve shows the phenomenological results
using $r_0 = 0.467\,\mbox{fm}$, a result we estimated previously,
see eg \cite{alikhan06a}. (This, of course, has the effect of making
our pseudoscalar masses larger and box size smaller in
table~\ref{table_status}.) It would seem that the strict applicability
of $\chi$PT is restricted to a rather narrow region
$r_{0c}m^{SS}_{ps} \lsim 1$; the choice of the scale is also rather delicate.
Another issue are possible finite-size effects, which we are planning to
investigate later.

Finally we note values of $f_{\pi^+} = 77(4)\,\mbox{MeV}$,
$f_{K^+} = 93(1)\,\mbox{MeV}$ for $r_0 = 0.5\,\mbox{fm}$ and
$f_{\pi^+} = 82(5)\,\mbox{MeV}$,
$f_{K^+} = 98(2)\,\mbox{MeV}$ for $r_0 = 0.467\,\mbox{fm}$.
Clearly using $f_{K^+}$ to set the scale would, at present, make the
lattice finer. The dimensionless ratio $f_{K^+} / f_{\pi^+}$
is $\sim 1.21$, $1.19$ for $r_0 = 0.5\,\mbox{fm}$, $0.467\,\mbox{fm}$
respectively, to be compared with $(f_{K^+} / f_{\pi^+})_{expt} \sim 1.223$.


\section*{Acknowledgements}


The numerical calculations have been performed on the Hitachi SR8000 at
LRZ (Munich), on the Cray T3E at EPCC (Edinburgh)
\cite{allton01a}, on the Cray T3E at NIC (J\"ulich) and ZIB (Berlin),
as well as on the APEmille and APEnext at DESY (Zeuthen),
while configurations at the smallest three pion
masses have been generated on the BlueGeneLs at NIC/J\"ulich, EPCC
at Edinburgh and KEK at Tsukuba by the Kanazawa group as part of the
DIK research programme. We thank all institutions.
This work has been supported in part by
the EU Integrated Infrastructure Initiative Hadron Physics (I3HP) under
contract RII3-CT-2004-506078 and by the DFG under contract
FOR 465 (Forschergruppe Gitter-Hadronen-Ph\"anomenologie).
We would also like to thank A.~C. Irving for providing updated results
for $r^S_0/a$ prior to publication.



\end{document}